\documentclass[a4paper,aps,twocolumn,nofootinbib]{revtex4}
\RequirePackage[colorlinks,hyperindex]{hyperref}
\RequirePackage[english]{babel}
\RequirePackage[latin1]{inputenc}
\RequirePackage[T1]{fontenc}
\RequirePackage{mathrsfs}
\RequirePackage{amsmath}
\RequirePackage{amssymb}
\RequirePackage{amsbsy}
\RequirePackage{color}
\RequirePackage{bm}
\hypersetup{colorlinks=true,breaklinks=true,urlcolor=blue,linkcolor=red}
\begin{document}
\title{\bf{Black Hole singularity avoidance by the Higgs scalar field}}
\author{Luca Fabbri}
\affiliation{DIME Sez. Metodi e Modelli Matematici, Universit\`{a} di Genova, 
Via all'Opera Pia 15, 16145 Genova ITALY}
\date{\today}
\begin{abstract}
Einstein gravitation is known to give rise to the formation of singularities at high densities unless the dominant energy condition is made invalid by the occurrence of new physics: we show that such a new physics can be the already present Higgs sector of the standard model of particle physics.
\end{abstract}
\maketitle
\section{Introduction}
In 2016, with the detection of the gravitational waves, we have witnessed the last and most impressive experimental confirmation of Einstein gravity; in 2012, with the discovery of the Higgs scalar, we have had the final and most important experimental confirmation of the validity of the standard model of particle physics. But despite all the success that the two theories have separately, nevertheless they still suffer the lack of a combined framework.

Even worse is that, theoretically, Einstein gravitation appears to be affected by the Penrose-Hawking theorem, stating that if the dominant energy condition holds then the high density gravitationally-induced formation of singularities is inevitable; in turn this means that the avoidance of singularity formation due to gravity can be done only by invalidating the condition on the energy, and thus what we would have to do would be to modify the structure of the energy tensor by allowing different contributions of matter distribution: this has been attempted in the past several times in \cite{kop, taf, tra, s-h, ke, Inomata:1976wi, Magueijo:2012ug, Khriplovich:2013tqa, Alexander:2014eva, Buchbinder:1985ym, Carroll:1994dq, Fabbri:2017rjf} by employing a variety of spinning fields, and it is a general feeling that, if not spin, at least some new type of matter field should enter the game in question. Along this line, it is counter-intuitive to think that the trick might be done by a field that is already known, and which does not even take into account spin, such as the Higgs scalar \cite{Fabbri:2009ta}; furthermore, it seems unlikely that the standard model contain the solution for a problem pertaining to gravity, since the two appear to have little in common. Just the same, Higgs physics can change the energy tensor in its structure, so to make the dominant energy condition invalid, and therefore to have the gravitationally-induced singularity avoided.
\section{Higgs sector of standard model with Einstein gravity}
As it is known, the standard model (SM) is the model built on the local $U(1)\!\times\!SU(2)$ symmetry group enriched with a mechanism of spontaneous symmetry breaking as induced by the Higgs scalar quartic potential: having the break-down of the symmetry, we can choose the so-called unitary gauge in which, after diagonalization of the mass matrix, we obtain the final Lagrangian, as it is presented in common textbooks; here we will not follow the usual textbook procedure of presenting the Lagrangian, but we will follow the different path of giving what follows from the Lagrangian via the variational approach, not just because it is an alternative way to present the SM, but also because there is more information in the field equations.

The entire set of SM field equations is in \cite{Fabbri:2009ta}, and again we have no need to represent it here in full; the sector in which we are interested is the one concerning the Higgs scalar, together with fermions, in Einstein gravity, and so we are going to consider the field equations in the limit in which all gauge fields are neglected, and where massless fields are left out: hence, the Higgs-electron-gravity part is described in terms of the field equations given by
\begin{eqnarray}
&\nabla^{2}H\!+\!m_{H}^{2}\left(\frac{H^{2}+3vH+2v^{2}}{2v^{2}}\right)H
\!+\!\frac{m_{e}}{2v}\overline{e}e\!=\!0
\label{Higgs}
\end{eqnarray}
and
\begin{eqnarray}
&i\boldsymbol{\gamma}^{\mu}\boldsymbol{\nabla}_{\mu}e\!-\!m_{e}\frac{H}{v}e\!-\!m_{e}e\!=\!0
\label{electron}
\end{eqnarray}
with
\begin{eqnarray}
\nonumber
&R_{\alpha\mu}\!-\!\frac{1}{2}g_{\alpha\mu}R\!=\!\frac{i}{8} (\overline{e}\boldsymbol{\gamma}_{\alpha}\boldsymbol{\nabla}_{\mu}e
\!-\!\boldsymbol{\nabla}_{\mu}\overline{e}\boldsymbol{\gamma}_{\alpha}e+\\
\nonumber
&+\overline{e}\boldsymbol{\gamma}_{\mu}\boldsymbol{\nabla}_{\alpha}e
\!-\!\boldsymbol{\nabla}_{\alpha}\overline{e}\boldsymbol{\gamma}_{\mu}e)+\\
\nonumber
&+(\nabla_{\mu}H \nabla_{\alpha}H\!-\!\frac{1}{2}\nabla_{\rho}H\nabla^{\rho}Hg_{\alpha\mu})+\\
&+\frac{1}{2}m_{H}^{2}H^{2}\left(\frac{H+2v}{2v}\right)^{2}g_{\alpha\mu}
\!-\!\frac{1}{8}v^{2}m_{H}^{2}g_{\alpha\mu}
\label{gravity}
\end{eqnarray}
in which $H$ is the Higgs scalar, $e$ is the electronic fermion field, $g_{\alpha\mu}$ is the metric of the space-time and $\boldsymbol{\gamma}^{\mu}$ are the Clifford matrices given in tetradic form. Notice that the covariant derivative applied to the scalar $\nabla_{\mu}H$ is written in a covariant notation but could simply be written as the partial derivative while the covariant derivative applied to the fermion $\boldsymbol{\nabla}_{\mu}e$ is the spinorial covariant derivative written in terms of the spin connection arising from the non-trivial metric of the space-time itself, and $R_{\alpha\mu}$ is the Ricci scalar; and finally $v$ is the non-trivial Higgs vacuum with $m_{H}$ and $m_{e}$ Higgs and electron masses. Notice that the Higgs scalar vacuum gives indeed rise to the negative cosmological constant term that entails the problematic contribution to the cosmological constant problem.

Once the gravitational field equations are provided, the Penrose-Hawking singularity theorem states that if
\begin{eqnarray}
&R^{\rho\sigma}u_{\rho}u_{\sigma}\!>\!0
\label{dec}
\end{eqnarray}
holds for some time-like vector $u^{\alpha}$ then singularities form inevitably \cite{Fabbri:2017rjf}. The (\ref{dec}) is the dominant energy condition.

To see if such a dominant energy condition (\ref{dec}) is valid, we need to write the gravitational field equations in the form in which the Ricci tensor is singled out: contracting the gravitational field equations (\ref{gravity}) gives
\begin{eqnarray}
\nonumber
&-R\!=\!\frac{1}{2}m_{e}\frac{H}{v}\overline{e}e\!+\!\frac{1}{2}m_{e}\overline{e}e-\\
&-\nabla_{\rho}H\nabla^{\rho}H\!+\!2m_{H}^{2}H^{2}\left(\frac{H+2v}{2v}\right)^{2}
\!-\!\frac{1}{2}v^{2}m_{H}^{2}
\end{eqnarray}
where we have used the fermion field equations and which can be plugged back into (\ref{gravity}) to give the final
\begin{eqnarray}
\nonumber
&R_{\alpha\mu}\!=\!\frac{i}{8} (\overline{e}\boldsymbol{\gamma}_{\alpha}\boldsymbol{\nabla}_{\mu}e
\!-\!\boldsymbol{\nabla}_{\mu}\overline{e}\boldsymbol{\gamma}_{\alpha}e+\\
\nonumber
&+\overline{e}\boldsymbol{\gamma}_{\mu}\boldsymbol{\nabla}_{\alpha}e
\!-\!\boldsymbol{\nabla}_{\alpha}\overline{e}\boldsymbol{\gamma}_{\mu}e)
\!+\!\nabla_{\mu}H \nabla_{\alpha}H-\\
\nonumber
&-\frac{1}{4}m_{e}\frac{H}{v}\overline{e}eg_{\alpha\mu}
\!-\!\frac{1}{4}m_{e}\overline{e}eg_{\alpha\mu}-\\
&-\frac{1}{2}m_{H}^{2}H^{2}\left(\frac{H+2v}{2v}\right)^{2}g_{\alpha\mu}
\!+\!\frac{1}{8}v^{2}m_{H}^{2}g_{\alpha\mu}
\end{eqnarray}
which can now be inserted into the dominant energy condition for final evaluation. After this (\ref{dec}) becomes
\begin{eqnarray}
\nonumber
&[\frac{i}{4}(\overline{e}\boldsymbol{\gamma}_{\alpha}\boldsymbol{\nabla}_{\mu}e
\!-\!\boldsymbol{\nabla}_{\mu}\overline{e}\boldsymbol{\gamma}_{\alpha}e)
\!+\!\nabla_{\mu}H \nabla_{\alpha}H-\\
\nonumber
&-\frac{1}{4}m_{e}\frac{H}{v}\overline{e}eg_{\alpha\mu}
\!-\!\frac{1}{4}m_{e}\overline{e}eg_{\alpha\mu}-\\
&-\frac{1}{2}m_{H}^{2}H^{2}\left(\frac{H+2v}{2v}\right)^{2}g_{\alpha\mu}
\!+\!\frac{1}{8}v^{2}m_{H}^{2}g_{\alpha\mu}]u^{\alpha}u^{\mu}\!>\!0
\label{DEC}
\end{eqnarray}
and this is the condition that we are next going to discuss.
\section{Effective approximation and comoving frame}
In principle the discussion would rely on finding some exact solution and see whether on shell (\ref{DEC}) is valid or not.

The actual discussion does not benefit from the leisure of exact solutions, and thus an alternative study must be performed; because the Higgs scalar is massive, the effective approximation can always be done and this allows the Higgs scalar degree of freedom to be integrated away.

The effective approximation is implemented by requiring that within the Higgs field equation (\ref{Higgs}) the dynamical term be irrelevant compared to the mass term as
\begin{eqnarray}
&m_{H}^{2}\left(\frac{H^{2}+3vH+2v^{2}}{2v^{2}}\right)H
\!+\!\frac{m_{e}}{2v}\overline{e}e\!\approx\!0
\end{eqnarray}
which in the limiting case of high densities further approximates down to the easier
\begin{eqnarray}
&H^{3}\!+\!\frac{vm_{e}}{m_{H}^{2}}\overline{e}e\!\approx\!0
\label{H}
\end{eqnarray}
allowing to integrate the Higgs scalar degree of freedom.

When equation (\ref{H}) is plugged into (\ref{DEC}) we obtain
\begin{eqnarray}
&\!\!\!\!\!\!\!\![\frac{i}{2}(\overline{e}\boldsymbol{\gamma}_{\alpha}\boldsymbol{\nabla}_{\mu}e
\!-\!\boldsymbol{\nabla}_{\mu}\overline{e}\boldsymbol{\gamma}_{\alpha}e)
\!+\!\frac{1}{4}\!\left(\frac{m_{e}\overline{e}e}{\sqrt{vm_{H}}}\right)^{\frac{4}{3}}
\!\!g_{\alpha\mu}]u^{\alpha}u^{\mu}\!>\!0
\label{aux}
\end{eqnarray}
in which we have kept only the most relevant terms.

When instead equation (\ref{H}) is plugged into (\ref{electron}) and we multiply the result by the conjugate spinor we get
\begin{eqnarray}
&i\overline{e}\boldsymbol{\gamma}^{\mu}\boldsymbol{\nabla}_{\mu}e
\!+\!\left(\frac{m_{e}\overline{e}e}{\sqrt{vm_{H}}}\right)^{\frac{4}{3}}\!=\!0
\end{eqnarray}
where again we have kept only the most relevant terms and which can be split in space and time parts as
\begin{eqnarray}
&i\overline{e}\boldsymbol{\gamma}^{t}\boldsymbol{\nabla}_{t}e\!=\!
-i\overline{e}\vec{\boldsymbol{\gamma}}\!\cdot\!\vec{\boldsymbol{\nabla}}e
\!-\!\left(\frac{m_{e}\overline{e}e}{\sqrt{vm_{H}}}\right)^{\frac{4}{3}}
\label{auxiliary}
\end{eqnarray}
in a form that will be needed to evaluate the time derivative of the fermion field in the dominant energy condition.

In fact, time-like vectors can always be boosted into a comoving frame, and there (\ref{aux}) takes the form
\begin{eqnarray}
&\frac{i}{2}(\overline{e}\boldsymbol{\gamma}_{t}\boldsymbol{\nabla}_{t}e
\!-\!\boldsymbol{\nabla}_{t}\overline{e}\boldsymbol{\gamma}_{t}e)
\!+\!\frac{1}{4}\!\left(\frac{m_{e}\overline{e}e}{\sqrt{vm_{H}}}\right)^{\frac{4}{3}}\!>\!0
\end{eqnarray}
which with (\ref{auxiliary}) becomes
\begin{eqnarray}
&-\frac{i}{2}(\overline{e}\vec{\boldsymbol{\gamma}}\!\cdot\!\vec{\boldsymbol{\nabla}}e
\!-\!\vec{\boldsymbol{\nabla}}\overline{e}\!\cdot\!\vec{\boldsymbol{\gamma}}e)
\!-\frac{3}{4}\!\left(\frac{m_{e}\overline{e}e}{\sqrt{vm_{H}}}\right)^{\frac{4}{3}}\!>\!0
\label{DECfinal}
\end{eqnarray}
as the dominant energy condition that we have to discuss.
\section{The scaling behaviour}
To discuss the dominant energy condition (\ref{DECfinal}) we may start from considering that in absence of the Higgs sector the only term that remains is the dynamical term
\begin{eqnarray}
&K\!=\!-\frac{i}{2}(\overline{e}\vec{\boldsymbol{\gamma}}\!\cdot\!\vec{\boldsymbol{\nabla}}e
\!-\!\vec{\boldsymbol{\nabla}}\overline{e}\!\cdot\!\vec{\boldsymbol{\gamma}}e)
\end{eqnarray}
with $K$ kinetic energy, always positive, hence recovering the old Penrose-Hawking result. But from a more general perspective, when this Penrose-Hawking analysis is done accounting for the Higgs sector, we get
\begin{eqnarray}
&K\!-\frac{3}{4}\!\left(\frac{m_{e}\overline{e}e}{\sqrt{vm_{H}}}\right)^{\frac{4}{3}}\!>\!0
\label{final}
\end{eqnarray}
displaying a remarkable difference: indeed, as an analysis of the scaling behaviour indicates, the kinetic energy term cannot suppress the Higgs-induced fermionic term.

In the evaluation of the dominant energy condition in this general case, both contributions are to be kept and the two have opposite effects, so that (\ref{final}) no longer holds as a necessity: in case the kinetic energy dominates, some singularities could still form; but when the Higgs-induced fermionic term dominates, singularities are avoided.

Moreover, cases with $K\!=\!0$ but $\overline{e}e\!\neq\!0$ entail unavoidable failure of (\ref{final}), and singularities become impossible in general; these situations are for instance those in which the fermionic fields can form condensed states.

We shall discuss one further example next.
\section{Some comments}
We have seen that in presence of the Higgs sector the dominant energy condition is given by (\ref{final}) with $K$ being the kinetic energy, always positive, while the other term is the Higgs contribution proportional to the fermionic density, always negative: hence, (\ref{final}) does not hold in all cases where the former term is irrelevant compared to the latter term, and singularities are avoided; and (\ref{final}) can not hold in all cases in which $K\!=\!0$ but $\overline{e}e\!\neq\!0$ and there singularities are impossible. Cases of zero kinetic energy but non-zero fermion density are those given by cold and dense distributions of matter, like condensates; however, this is even more so for extreme situations such as those that are encountered in a Black Hole. As a consequence we should expect Black Holes to display no singularity.

Notice that it is rather reasonable that the Higgs scalar produce an interaction between fermions with an attractive character: indeed, this can be seen as the remnant of the attractive interaction between left-handed and right-handed chiral parts that the Higgs needs to exert in order to give mass to the fermion. In fact, it is the interplay between chiral parts what gives mass to the fermion field.

Notice also that it may look paradoxical that this attractive interaction be responsible for a counter-balance of the gravitational force, but there is quite a simple explanation for this phenomenon in the Einsteinian theory of gravitation: in Einstein theory of gravity the gravitational force is thought to be encoded within the curvature of the space-time, which in general gives rise to gravitational pull because matter distributions in general have a positive energy. However, an attractive interaction, entailing a potential well, dispenses negative contributions inside the total energy, and in situations in which these are the most relevant of the contributions the energy reverts to negative, and the space-time curvature relaxes.

As we have said, this may look rather counter-intuitive, but it is a natural consequence of the Einsteinian gravity, where the gravitational field is the space-time curvature and as such it is susceptible of being positive as well as being negative; and similarly it is the natural consequence of the presence of the Higgs sector, and the fact that it does give rise to a generally attractive interaction.

It is also important to notice that Einstein gravity with the Higgs scalar but with no fermion field would still be insufficient to have the singularity avoidance.
\section{Conclusion}
In this paper, we have considered the SM after symmetry breaking, and isolating only the Higgs-electron sector, we have exhibited the field equations, and we have considered the Einstein gravitational field equations sourced by the energy tensor of this scalar-fermion sector; to discuss the influence of the scalar-fermion sector in the dominant energy condition, we have assumed the effective approximation for the Higgs field and we have boosted into the frame comoving with the electronic field: we found that the dominant energy condition in this case accounted for the kinetic term, always positive, plus the fermion-Higgs term, always negative. This makes the dominant energy condition fail to be valid as a necessity; we have recalled cases where the dominant energy condition does not hold, and cases in which the dominant energy condition cannot hold at all. Therefore, singularity formation is sometimes avoided, and in some situations it is totally impossible.

We have stated that such circumstances are those that we have for example in Black Holes: because Black Holes are gravitational structures constituted by fermion fields in interaction through the Higgs scalar, being very cold and highly dense they are precisely the case of object for which the dominant energy condition cannot be valid, so that the gravitational singularity formation is impossible.

As a consequence of this fact, we have that Black Holes cannot be gravitationally singular as a mathematical consequence of the theorem of Penrose and Hawking.

However, they may still be gravitationally singular as a consequence of other mechanisms: as a start, all these results may change considerably without the effective approximation on the Higgs field; additionally, there may be differences of dramatic importance when employing field equations before symmetry breaking in SM. Finally, Black Holes may still form gravitational singularities in Einstein gravity's extended versions or its alternatives.

For example, the lack of an effective approximation, or even worse the thorough lack of the symmetry breaking, makes it unreasonable to have these results stretched as to include the gravitational singularities we could find in Big Bang cosmologies, and therefore we cannot have any confidence in arguing that the Big Bang itself could turn out to be free of any gravitational singularity.

There is a similar impossibility in having these results enlarged to Einstein gravity's extensions or alternatives, where the different structure of the background requires a complete revision of the singularity theorems.

In these cases, Black Holes or cosmologies free of any singularity may still be possible \cite{Zhang:2018qdk}.

\end{document}